\documentclass[twocolumn,aps,prl,showpacs,superscriptaddress]{revtex4}
\usepackage{graphicx}
\begin{document}

\author{Arne Brataas}
\affiliation{Department of Physics, Norwegian University of Science
and Technology, N-7491 Trondheim, Norway}

\author{Gergely Zar\'and}
\affiliation{Theoretical Physics Department, Budapest University of 
Technology and Economics, H-1521 Hungary}

\author{Yaroslav Tserkovnyak}
\affiliation{Lyman Laboratory of Physics, Harvard University,
Cambridge, Massachusetts 02138}

\author{Gerrit E. W. Bauer}
\affiliation{Department of NanoScience, Delft University of
Technology, 2628 CJ Delft, The Netherlands}

\title{Magnetoelectronic Spin Echo}

\begin{abstract}
We predict a spin echo in electron transport through layered
ferromagnetic-normal-ferromagnetic metal structures: whereas a spin current polarized perpendicular to the magnetization direction decays when traversing a single homogeneous ferromagnet on the scale of the ferromagnetic spin-coherence length, it partially reappears by adding a second identical but antiparallel ferromagnet. This re-entrant transverse spin current resembles the spin-echo effect in
the magnetization of nuclei under pulsed excitations.  We propose an experimental setup to measure the spin echo.
\end{abstract}
\date{\today}
\maketitle


The spin echo \cite{Hahn:pr50} is a remarkable phenomenon:
Subjecting a magnetized spin system to two rf pulses in succession with a time delay $\tau$, the initially destroyed magnetic order reappears at time $\tau$ after the second pulse seemingly out of nowhere. Hahn \cite{Hahn:pr50} explained the spin echo with the following simple picture: Imagine runners at a race track starting at the same time. Variations in their speed spread out the platoon. However, if we order the runners to suddenly turn around at time $t=\tau$ (rf pulse) and start running with the same speed in the opposite direction, they all meet again at $t=2 \tau$.

In this Letter we propose a novel kind of spin echo in magnetic nanoscale spin valves, i.e., ferromagnetic-normal-ferromagnetic (F/N/F) layered metal systems with antiparallel magnetization alignment. Spin valves are the building block of current-perpendicular-to-plane giant-magnetoresistance devices \cite{Gijs:adv97}. Furthermore, in spin valves the reversal of magnetizations by a current bias has been observed \cite{Rev}, which is caused by the spin torque \cite{torque}, i.e., absorption of spin-current components transversely polarized to the magnetization direction.

Before delving into the theory, let us explain our main idea. Consider a spin current which is injected into a ferromagnetic metal with a polarization perpendicular to its magnetization. The transverse spin state is not an eigenstate of the ferromagnet, but a coherent linear combination of the majority and minority spin eigenstates with different Fermi wave vectors $k_{F\uparrow}$ and $k_{F\downarrow}$. The spin therefore precesses on a length scale depending on the perpendicular component of the wave-vector difference at the Fermi energy, which depends on the angle of incidence. In elemental ferromagnets like Co, Ni and Fe, a large number of modes with different spin precession lengths in the ferromagnet contribute to the total current. Their destructive interference leads to a {\em loss of transverse spin current} inside the ferromagnet on a length scale of $\lambda_{\text{sc}}=\pi/|k_{F\uparrow}-k_{F\downarrow}|$ \cite{Brataas:prl00,Stiles:prb02} (the so-called transverse spin-coherence length), which for typical transition metals is only a few Angstroms. The lost angular momentum is transferred to the ferromagnetic condensate, which thus experiences a torque that can lead to a magnetization reversal, as predicted by Slonczewski and Berger \cite{Rev}. In this Letter we investigate (anti)symmetric F/N/F spin valves, in which the magnetization of the second ferromagnet points into the opposite direction of the first one. In such a structure the spin precession in the first ferromagnet is time reversed in the second ferromagnet, the lost transverse spin current is recovered, and the total spin torque on the magnetizations, i.e., the absorbed spin angular momentum, is significantly reduced. Each propagating mode experiences a real reversal on the spin coordinate on traversing the spin valve. Consequently a transverse spin current can propagate through the (anti)symmetric F/N/F spin valve. This is the magnetoelectronic
spin echo.

Consider a section of a magnetoelectronic circuit as depicted in Fig. 1. The scattering region in the center is connected to a normal metal on the left ($i=1$) and 
to a normal metal on the right ($i=2$). Let us introduce the non-equilibrium distribution function $\hat{f}^{(i)}(\epsilon)$ for electrons at energy $\epsilon$, which is a $2\times 2$ matrix in Pauli spin space. The chemical potential matrix $\hat{\mu}^{(i)}= \int d\epsilon \hat{f}^{(i)}(\epsilon)$ contains a scalar charge contribution $\mu^{(i)}_0=\text{Tr} \left[\hat{\mu}^{(i)}\right]/2$ and a vector spin
contribution usually called spin accumulation ${\boldmath{\text{$\mu$}}}=\text{Tr} \left[ \hat{\mu}^{(i)} \boldmath{\text{$\sigma$}} \right]/2$, where
$\boldmath{\text{$\sigma$}}$ is the vector of Pauli matrices. Note that with these definitions the equilibrium values are $\mu_0^{(1)}=\mu_0^{(2)}$ and ${\boldmath{\text{$\mu$}}}^{(1,2)}=0$. 

\begin{figure}[tbp] 
\includegraphics[width=0.9\linewidth]{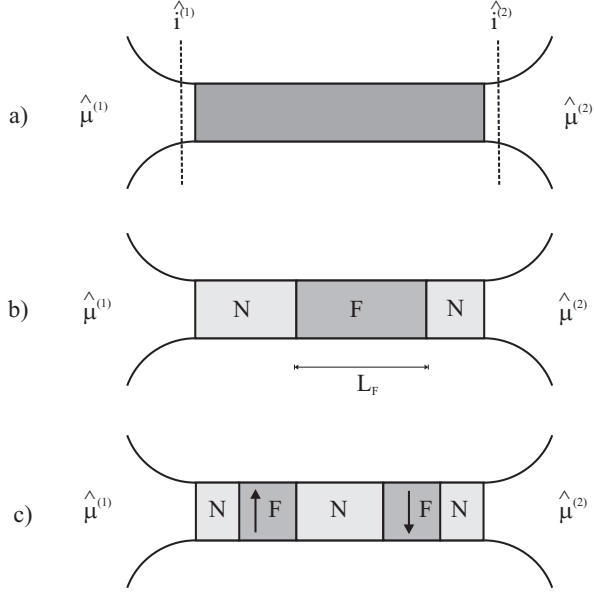} 
\caption{a) The scattering region couples the left (1) and
right (2) normal-metal nodes with nonequilibrium chemical potentials
$\hat{\mu}^{(1)}$ and $\hat{\mu}^{(2)}$.
b) The scattering region consists of one uniform ferromagnet and normal metals. c) The scattering region consists of two
identical antiferromagnetically-aligned ferromagnets separated by a normal metal.
A partial electron spin echo can be seen in transport in this configuration.} \label{f1} 
\end{figure} 

We assume that in the left and right normal metals in Fig. 1 spin accumulations are excited by external circuits to be discussed later. We are interested in the $2\times 2$ matrix currents $\hat{i}^{(i)}$
at the left- and the right-hand side of the scattering region which are induced by these spin accumulations. When spin currents and ferromagnetic magnetization directions are not collinear, $\hat{i}^{(1)} \ne \hat{i}^{(2)}$, i.e., spin current is not conserved. The currents between two normal metals with an out-of-equilibrium spin and charge accumulation can be found either with the Keldysh technique following Ref.~\cite{Brataas:prl00} or by using the Landauer-B\"{u}ttiker formalism \cite{Tserkovnyak:prb01}. We find for the left (1) and the right side (2) the following expressions for the matrix currents
\begin{eqnarray}
\hat{\imath}^{(1)} &=&\frac{1}{2}\sum_{nm}\left[ \hat{t}_{nm}\hat{\mu}^{(2)} \hat {t}
_{nm}^{\ast}-\hat{\mu}^{(1)}+\hat {r}_{nm}\hat {\mu}^{(1)} 
\hat {r}_{nm}^{\ast }\right] \, ,
\\ 
\hat{\imath}^{(2)}&=&\frac{1}{2}\sum_{nm}\left[\hat {t}_{nm}^{\prime }\hat{\mu}^{(1)} \hat {t}_{nm}^{\prime\ast}-\hat{\mu}^{(2)}+\hat {r}_{nm}^{\prime }\hat{\mu}^{(2)} \hat{r}_{nm}^{\prime\ast}\right]  \, ,
\end{eqnarray}
where $\hat{r}_{n,m}$, $\hat{t}_{n,m}$, $\hat{r}^{\prime }_{n,m}$, and
$\hat{t}^{\prime }_{n,m}$ are reflection and transmission matrices of the scattering region both in the space spanned by the transverse channels (momentum labeled by $n$ and $m$) and the spin space (denoted by the hats).

For a specific scattering region with fixed ferromagnetic magnetization directions parallel or antiparallel to the spin-quantization axis ($z$ direction) and disregarding spin-flip scattering, the transmission matrices are spin diagonal, $(\hat{t}_{nm})_{s,s'} = \delta_{s,s'} t_{nm}^s$, where $t_{nm}^{s}$ is the transmission coefficient for an electron with
spin $s$ ($s=\uparrow$, parallel, and $\downarrow$, antiparallel). Same holds for the transmission coefficients $t_{nm}^{\prime s}$ as well as the reflection coefficients $r_{nm}^{s}$ and $r_{nm}^{\prime s}$.

The spin and charge currents through the scattering region depend on both the charge and the spin chemical potentials. In order to simplify the discussion, we assume in the following that the normal metals are spin but not charge biased, $\mu^{(1)}_0=\mu^{(2)}_0$. This is a perfectly realistic situation, as outlined in Refs.~\cite{Brataas:prb02, Tserkovnyak:prl02}, and does not imply any loss of generality for our purposes. We first compute the spin current ${\bf I}^{(2)}=\text{Tr} \left[ \hat{i}^{(2)} \boldmath{\text{$\sigma$}} \right]$ on the right of the scattering region and find its components to be:
\begin{eqnarray}
I_{x}^{(2)} &=& \left[  g_{R}^{t'} \mu_{x}^{(1)} +  g_{I}^{t'} \mu_{y}^{(1)} \right] - \left[ g_{R}^{r'} \mu_{x}^{(2)} + g_{I}^{r'} \mu_{y}^{(2)} \right] \, ,
\nonumber \\
I_{y}^{(2)} &=&  \left[ g_{R}^{t'} \mu_{y}^{(1)} - g_{I}^{t'} \mu_{x}^{(1)} \right] - \left[ g_R^{r'} \mu_{y}^{(2)} - g_I^{r'} \mu_{x}^{(2)} \right] \, ,
\nonumber
 \\
I_{z}^{(2)} &=& g \left[ \mu_{z}^{(1)} - \mu_{z}^{(2)} \right] \, ,
\label{eq:I_right}
\end{eqnarray}
where the conductances associated with transmission and reflection are $g^{t'} = \sum_{nm}t_{nm}^{\prime \uparrow
}t_{nm}^{\prime \downarrow \ast }$, 
$ g^{r'}  = \sum_{nm}\left[ \delta_{nm}-r_{nm}^{\prime \uparrow }
r_{nm}^{\prime \downarrow \ast }\right]$, 
$g =(1/2)\sum_{nm} \left[ \left| t_{nm}^{\prime \uparrow } \right| ^{2} 
+ \left|t_{nm}^{\prime \downarrow }\right|^2 \right]$, and the subscripts $R$ and 
$I$ refer to real and imaginary parts, respectively. Similarly, the current on the left of the scattering region ${\bf I}^{(1)}$ is expressed as Eq.\ (\ref{eq:I_right}) with the conductances $g^{t}$ and $g^{r}$ denoting the substitutions $t' \rightarrow t$ and $r' \rightarrow r$ and with the substitutions $1 \leftrightarrow 2$. 

Let us now discuss the physical content of Eq.~(\ref{eq:I_right}). First, the longitudinal component of the spin current (i.e., the component of the spin current collinear to the magnetization in the scattering  region, which is the $z$ direction) is conserved, as expected, $I^{(1)}_z + I^{(2)}_z = 0$. Second, if the scattering region would be a normal metal, the transmission and reflection coefficients are spin independent and consequently the $x$ and $y$ components of the spin current are conserved as well. When the magnetization directions in the scattering region are noncollinear with the spin accumulations, the transverse spin current is not necessarily conserved, so that in general $I^{(1)}_x + I^{(2)}_x \ne 0$, $I^{(1)}_y +I^{(2)}_y \ne 0$. 
As explained above, for a single ferromagnetic layer thicker than the ferromagnetic spin-coherence length $\lambda_{\text{sc}}$, we know that $g^{t}$ and $g^{t'}$ vanish rapidly. The transverse spin current is then exclusively determined by the (reflection) spin-mixing conductance $g^r$ \cite{Brataas:prl00}. 
For high-density metallic systems, the phases of $r^\uparrow$ and $r^\downarrow$ are large and uncorrelated, and therefore 
$g^{r} \approx g_{\text{S}} \approx g^{r'}$, where $g_{\text{S}}$ is the Sharvin conductance given by the number of propagating channels in the normal-metal leads. The conventional conductance $g$ for the longitudinal spin component is determined by the transmission probability for spin-up and spin-down electrons. The spin currents are then simply 
$I_z^{(2)}=g(\mu_z^{(1)}-\mu_z^{(2)})=-I_z^{(1)}$, and 
$I_{x,y}^{(1,2)} = -g_{\text{S}} \mu_{x,y}^{(1,2)}$. The latter expression 
represents the loss of spin current that is directly proportional 
to the spin torque \cite{Xia:prb02}.

Let us now turn to the main subject of this paper, the F/N/F spin valve
in Fig.~1c. The lengths of the left and right leads are irrelevant for the transport properties, and we may define a scattering matrix for the left half of the scattering region (consisting of the left normal-metal lead with a length equal half of the normal metal spacer, the left 
ferromagnet, and half of the normal metal spacer) and the similar right half. The total scattering matrix can be found by concatenation, most conveniently for a spin-quantization axis collinear to the magnetization. The (spin-dependent) transmission and reflection matrices for the entire scattering region then read
\begin{eqnarray}
t^{s} & = & t_2^{s} [1-r_1^{'s}r_2^{s}]^{-1} t_1^{s} \label{trans_concat} \\
r^{s} & = & r_1^{s} + t_1^{'s} r_2^{s} [1-r_1^{'s}r_2^{s}]^{-1} t_1^{s} \label{ref_concat}
\end{eqnarray}
where $s$ denotes the spin of the electron and subscripts 1 and 2 left and right regions of the scatterer, respectively. Similar expressions hold for $t'$ and $r'$. 

The spin valve is usually opaque for the transverse spin component. In order
to observe the spin echo, the device must be structurally {\em symmetric} and {\em clean}, with antiparallel magnetization configurations. 
In general, the transmission (\ref{trans_concat}) and reflection (\ref{ref_concat}) matrices are nondiagonal in the space spanned by the transverse waveguide modes.  
For clean systems with specular scattering, reflection and transmission matrices in Eqs.~(\ref{trans_concat}) and (\ref{ref_concat}) are diagonal in the transverse wave vector \cite{fermi}. Furthermore, for mirror-symmetric ferromagnetic layers and an antiparallel and symmetric F/N/F spin valve for a suitable gauge choice symmetry dictates that
$t_{1,ii}^{\uparrow}=t_{1,ii}^{'\uparrow}=t_{2,ii}^{\downarrow}=t_{2,ii}^{'\downarrow}$, 
$t_{1,ii}^{\downarrow}=t_{1,ii}^{'\downarrow}=t_{2,ii}^{\uparrow}=t_{2,ii}^{'\uparrow}$, 
$r_{1,ii}^{\uparrow}=r_{1,ii}^{'\uparrow}=r_{2,ii}^{\downarrow}=r_{2,ii}^{'\downarrow}$, 
and $r_{1,ii}^{\downarrow}=r_{1,ii}^{'\downarrow}=r_{2,ii}^{\uparrow}=r_{2,ii}^{'\uparrow}$, so that
\begin{equation}
t^{\uparrow} = t^{\downarrow} = t^{\downarrow}_{1} t_{1}^{\uparrow} 
[1-r_{1}^{\uparrow} r_{1}^{\downarrow}]^{-1}  \, .
\end{equation}
We find that $g^{t}_I=0$ and $g_I^{t'}=0$, but in contrast to the single ferromagnetic layer, $g_{R}^{t}$ and $g_{R}^{t'}$ are nonzero, which implies that a transverse spin current is permitted now.
This is the {\em spin echo} in electron transport: The transverse spin coherence apparently lost on traversing the first ferromagnet, reappears after passing through the second ferromagnet, since $g^{t}=g^{t'}=g$. On the other hand there is no connection between $r^{\uparrow}$ and $r^{\downarrow}$. Consequently, the reflection mixing conductance is approximately equal to the Sharvin conductance, $g^{r} \approx g_{\text{S}}$, and is not affected by the second ferromagnet. The current in the right lead is thus
\begin{eqnarray}
I_{\alpha}^{(2)} & = & g \mu_{\alpha}^{(1)} - g^{r} \mu_{\alpha}^{(2)}
\;, 
\phantom{mmm}(\alpha = x,\phantom{n}y)\;,
\label{Ixy2_echo} 
\\
I_z^{(2)} & = & g \left[\mu_{z}^{(1)} - \mu_{z}^{(2)} \right]
\end{eqnarray}
and the current in the left lead becomes
\begin{eqnarray}
I_{\alpha}^{(1)} & = & g \mu_{\alpha}^{(2)} - g^{r} \mu_{\alpha}^{(1)} 
\;, 
\phantom{mmm}(\alpha = x,\phantom{n}y)\;,
\label{Ixy1_echo} 
\\
%
%
I_z^{(1)} & = & g \left[\mu_{z}^{(2)} - \mu_{z}^{(1)} \right]\,.
\end{eqnarray}
The first terms on the right-hand side of Eqs.\ (\ref{Ixy2_echo}) and (\ref{Ixy1_echo}) provide a quantitative expression for our spin echo in clean systems. Note that, although a transverse spin current is allowed through the sample, it is {\em not conserved}, $I_{x}^{1} + I_{x}^{2}\ne 0$, 
and $I_{y}^{1} + I_{y}^{2}\ne 0$, as $ g^{r} \approx g_{\text{S}}>g$, and the total magnetization torque on the spin valve is not completely quenched. Whereas the transmission mixing conductance is now strongly enhanced, the reflection mixing conductance is essentially unmodified, as noted above, and continues to exert a torque on the spin valve. The transmission of the transverse spin current suppresses an important contribution to the torque present in single-layer ferromagnets \cite{Waintal:prb00,Stiles:prb02}, however.

In order to observe the spin echo, the transverse wave vector must be conserved, because otherwise the electron precession through the second ferromagnet is not exactly time-reversed compared to that of the first ferromagnet. Disorder such as impurity scattering, interface alloying, and layer thickness fluctuations is detrimental to the spin echo. The coherent propagator or transmission coefficient is exponentially damped on the length scale of the mean free path. This is not of much concern, because the mean free path in bulk materials can be much larger than the film thicknesses in multilayers. More critical is the interface quality, since monolayer fluctuations of the layer thickness can already significantly dephase the spin echo. One might therefore question the observability of the spin echo in state-of-the-art transition-metal structures in which transport is usually well described by the classical series-resistor model \cite{Gijs:adv97}. There are several reasons to be optimistic, however. For one, the nonlocal exchange coupling in magnetic multilayers is a robust quantum-interference effect routinely observed in multilayers, oscillating not only as a function of the normal- but also of the magnetic-layer thickness \cite{xcoup}. Furthermore, transport experiments with focused electrons, either by tunneling barriers \cite{suzuki} or by hot carriers \cite{buhrman} are indicative of specular scattering at high-quality interfaces. Indeed, conventional transport in high-density metallic structures consists of a large semiclassical background with relatively small quantum corrections \cite{Tsymbal:ssp01}. In contrast, the spin-echo signal vanishes in the semiclassical approximation. The suppression of the spin echo by disorder thus provides direct information about the degree of quantum interference in electron transport.

First-principles band-structure calculations can provide quantitative 
estimates for the magnitude of the spin echo \cite{Xia:prb02}. Such 
calculations can provide important information on the effect of 
interband scattering, impurity scattering at interfaces and bulk 
disorder on the suppression of the spin echo. We expect that a thickness difference between the two ferromagnetic layers would roughly suppress the spin echo as a single ferromagnetic layer of width equal the difference would suppress a transverse spin current. We thus estimate that the spin echo signal decays to 10-20 \% of its original value by one atomic layer mismatch, and that the decay is algebraic\cite{Stiles:prb02}. A small misalignement $\theta$ of the magnetizations is not detrimental to the spin echo since the difference in phase shifts traversing the ferromagnets is small, $\Delta \phi = k_F L_F [\cos(\theta) - 1]$ $\approx$ $k_F L_F \theta^2/2 > 2 \pi$, which gives $\theta^2 < 2 (\lambda_F / L_F)$. This is easily satisfied e.g. for $L_F = 20 \lambda_F$ resulting in $\theta < 18$ degrees.

\begin{figure}[tbp] 
\includegraphics[width=0.9\linewidth]{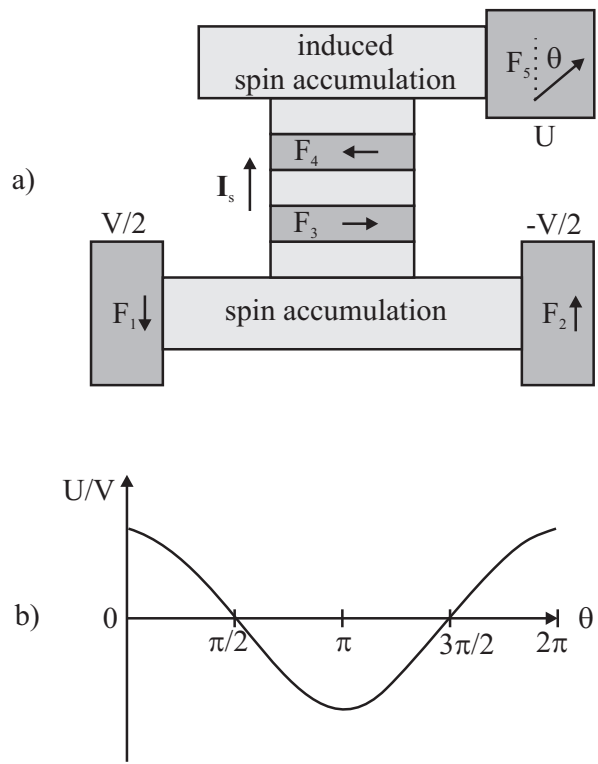} 
\caption{
Proposed experiment to observe the spin echo. Grey areas denote  
ferromagnetic layers. The spin echo is measured via the angular dependence of the potential $U(\theta)$ (or simply by its drop when $\theta$ is changed from $0$ to $\pi$).} 
\label{fig:experiment} 
\end{figure} 

We suggest a double-layer thin-film arrangement shown in Fig.~\ref{fig:experiment}a for experimental observation of the spin echo. The $F_3/N/F_4$ spin valve should be thinner than the mean free path for impurity scattering. A Co/Cu/Co structure with a copper-layer thickness corresponding to the first antiparallel-coupling energy minimum should be a good choice because the spin flip in these materials is weak. Besides, this system, in particular its fabrication, is thoroughly investigated. The transverse spin accumulation driving the spin current through layers $F_3$ and $F_4$ can be excited by a spin battery operated by ferromagnetic resonance \cite{Brataas:prb02} or by a current-biased antiparallel spin valve with ferromagnets $F_1$ and $F_2$ whose magnetizations are rotated by 90 degrees with respect to $F_3$ and $F_4$, analogously to the spin-torque transistor \cite{Bauer:apl03}. The biased ferromagnets $F_1$ and $F_2$ create a spin accumulation in the normal metal between them. This spin accumulation can traverse the $F_3$/$F_4$ spin valve only when conditions for the spin echo are fulfilled. In that case, an induced spin accumulation in the upper normal metal, and hence the spin echo, can be detected as a voltage depending on the magnetization angle $\theta$ of ferromagnet $F_5$ as sketched in Fig. 2b.

In conclusion, we predict a magnetoelectronic spin echo in spin valves. This is a sensitive measure of quantum coherence in metallic magnetic multilayers.

We would like to thank B.\ I.\ Halperin and B. J. van Wees for stimulating discussions. This work was supported in part 
by the NEDO International Joint Research Grant Program \textquotedblleft Nano-magnetoelectronics\textquotedblright, Hungarian Grants No. 
OTKA F030041, T038162, the EU Spintronics RTN HPRN-CT-2002-00302 and FOM. G.\ Z.\ is a Bolyai Fellow.

Note added: The supercurrent through anti-parallel spin valves predicted independently by Blanter and Hekking [cond-mat/0306706] can be interpreted as a manifestation of our spin echo.


\begin{thebibliography}{99}
\bibitem{fermi} Here it is implied that the outermost normal metals have simple Fermi surfaces (like Cu), but the principal arguments hold also for multiband normal metals.
\end{thebibliography}
\end{document}